\documentclass[preprint]{article}
\usepackage{graphicx}

\newcommand{\be}{\begin{equation}}
\newcommand{\ee}{\end{equation}}
\newcommand{\ba}{\begin{eqnarray}}
\newcommand{\ea}{\end{eqnarray}}

\begin{document}
\setlength{\textheight}{8.5 in}

\begin{titlepage}
\begin{center}
{\Large\bf {Conformal Bootstrap Analysis for Localization: Symplectic Case }}
\vskip 6mm
 Shinobu Hikami 

\vskip 5mm

Mathematical and Theoretical Physics Unit,
Okinawa Institute of Science and Technology Graduate University,
Okinawa, Onna 904-0495, Japan; hikami@oist.jp

\end{center}
\vskip 20mm
\begin{center}
{\bf Abstract}
\vskip 3mm
\end{center}
The localization phenomena due to the random potential scattering is widely discussed in the electron and photon systems. The theoretical approach by the nonlinear $\sigma$ model with the replica method  or with the supersymmetry has been investigated near two dimensions.  In this article, we consider the application of the  conformal bootstrap method
to the symplectic Anderson localization for the arbitrary dimensions.

\vskip 3mm
 \end{titlepage}
 \newpage
 \section{Introduction}
\vskip 2mm
The conformal bootstrap method has  a  powerful algorithm for obtaining very precise  critical exponents in  three dimensional Ising model. It has been applied to other interesting cases. The references may be found in a recent review article  \cite{Rychkov2018}, which includes the basic explanation of the conformal bootstrap method. The conceptual and technical explanation of the conformal bootstrap method may be found in this review. The successful cases are unitary cases, which require the unitary condition that the coefficients of the
expansion of the operator product expansion are positive. However, in the critical phenomena, there are non-unitary cases such that the Yang-Lee edge singularity and polymer case.
The conformal bootstrap method by the use of  small determinants has been applied for the non-unitary model of Yang-Lee edge singularity \cite{Gliozzi2013, Hikami2018a}.
The polymer and the branched polymer, which are also non-unitary,  are considered by the conformal bootstrap method \cite{Hikami2018b}.
The random systems are treated by the $N\to 0$ limit of $O(N)$ invariant models due to the replica method. The random magnetic field Ising  model (RFIM) is studied also by the conformal bootstrap  method \cite{Hikami2018c}.

In this article, we apply this non-unitary conformal bootstrap method to the localization problems. The Anderson localization has passed more than 50 years from the discovery, and has been applied to electron systems and photon systems \cite{Abrahams2010}.
The renormalization group method for the localization is based on the field theoretical model, known as the nonlinear $\sigma$ model with the limit $N\to 0$ of Grassmann manifold $O(2N)/O(N)\times O(N)$ or its non-compact version \cite{Hikami1981}. The isomorphic relation $O(-2N) = Sp(N)$ converts the orthogonal to the symplectic Grassmannian manifold.
The symplectic localization corresponds to  the limit of $p=N=0$ in $Sp(N)/Sp(p)\times Sp(N-p)= O(-2N)/O(-2p)\times O(-2N+2p)$, and we take the coupling constant $t$ of this nonlinear sigma model to $t\to -t$ from the reason of the non-compactness \cite{Hikami1980a,Hikami1980b}. In two dimensions, $\beta$-function is not asymptotic free, and similar to the Ising model, and it suggests the existence of the phase transition in two dimensions. There are many types of the nonlinear $\sigma$ models, defined on the symmetric spaces. The polymer case has critical phenomenon , in the replica limit $N\to 0$,  in two dimensions. The symplectic localization case, in the replica limit $N\to 0$, has a phase transition.
Although we assume the present analysis for the symplectic localization by $Sp(2N)/Sp(N)\times Sp(N)$, there remains a possibility  that the present bootstrap analysis may describes the localization of $Sp(N)/U(N)$  non-linear $\sigma$ model in the replica limit $N\to 0$ \cite{Hikami1981b}.

Among three different types of the Anderson localization, the symplectic case, where the time reversal symmetry is preserved but the symmetry of space inversion is broken, leads to 
the existence of the phase transition in two dimensions. The strong spin-orbit  case belongs to  the  symplectic case \cite{Hikami1980a}.
The numerical analysis of the  localization has precise predictions of the critical exponents due to the finite scaling \cite{Minakuchi,Minakuchi2,Asada2002}, and the Borel-Pad'e re-summation for the results of $\beta$-function has been proposed to obtain the exponents \cite{Ueoka2017}.
We specify the model, which satisfies the conditions of (\ref{condition}) in below,  and we call it as a localization model, in which  the density of state does not show the any singular behavior.

  Our study is a continuation of the previous works \cite{Hikami2018a,Hikami2018b,Hikami2018c}. These cases have characteristic properties that two scale dimensions are degenerated.  In the $N\to 0$ limit in $O(N)$ invariant system, it is important to notice that the crossover exponent $\hat\varphi$ of $O(N)$ vector model \cite{Hikami1974} becomes one in the $N\to 0$ limit, and hence two scaling dimensions $\Delta_\epsilon=D - \frac{1}{\nu}$ and $\Delta_T= D - \frac{\hat\varphi}{\nu}$ become degenerated, where $D$ is the space dimension and $\nu$ is the critical exponent of the correlation length. This degeneracy is a characteristic behavior of the polymer in any dimensions. For Yang-Lee edge singularity, similar degeneracy occurs as two scale dimensions $\Delta_\phi$ and $\Delta_\epsilon$. The notation of the scale dimension $\Delta_\phi$ is the scale dimension of the field $\phi$, which is defined in (\ref{Dphi}) by the critical exponent $\eta$. For the polymer case, we have used a blow up technique \cite{Hikami2018b,Hikami2018c}. The blow up means the resolution of the degeneracy by giving small difference between $\Delta_\epsilon$ and $\Delta_T$. The blow up may be similar to the resolution of the singularity in the algebraic geometry \cite{Hironaka1964, Nash1995}.

For the localization phenomena, the system becomes localized due to the random potential or random scattering. The diffusion coefficient becomes null when the localization occurs.
The diffusion propagator appears in two body Green function. The one body Green function, which imaginary part is the density of state, has no singularity. This means the critical exponent
$\beta$ is vanishing for the localization problem \cite{Hikami1981}. The exponent $\beta$ is given by the scaling law,
\be
\beta = \nu ( D - 2 + \eta)
\ee
Since the scale dimension $\Delta_\phi$ is
\be\label{Dphi}
\Delta_\phi= \frac{D- 2 + \eta}{2}
\ee
we have a condition of $\Delta_\phi=0$ for the localization in any space dimensions $D$. 

 Thus we have a set of the constraints for the scale dimensions in the localization as
 \be\label{condition}
 \Delta_\phi= 0, \hskip 3mm \Delta_T = \Delta_\epsilon
 \ee
 Some investigation has been done in the Yang-Lee edge singularity for the critical dimensions $D_c$, in which $\Delta_\phi=\Delta_\epsilon = \Delta_T= 0$ is realized \cite{Hikami2018a}, and it satisfies the constraint of (\ref{condition}). Therefore, the situation of this particular case in Yang-Lee edge singularity is analogous to the localization problem.
 
 The conformal bootstrap does not need the Lagrangian, and instead,  only the symmetrical constraints determine the scale dimensions. If the fusion rule is known, the conformal bootstrap method is easily applied ststematically. However, the precise fusion rule is not known for the localization problem.   For the localization, the usual method
  is starting from the nonlinear $\sigma$ model, and the perturbational expansion gives the $\beta$-function successively. There is a different method, which uses the scattering amplitude like Virasoro-Shapiro amplitude of the string theory, and it determines the renormalization group $\beta$ function \cite{Hikami1992}. The conformal bootstrap method is generally related  to 
  the four point scattering amplitude through AdS/CFT correspondence \cite{Witten1998}. In this paper, assuming we don't know the fusion rule for the operator product expansions, we simply use the approximation by the use of small numbers of determinants, which involve a limited number of the scale dimensions. We have obtained the values of the critical exponent $\nu$ which  roughly agrees with the numerical estimation by the finite size scaling method, and also we obtain the lower critical dimension $D_c=1.25$ and the upper critical dimension as $D_c=\infty$, which are consistent with the numerical result of  finite size scaling \cite{Asada2002,Ueoka2017}. For the case of polymer \cite{Hikami2018b}, the analysis of small size determinants provide a rather accurate value of the critical exponent $\nu$. Therefore, for the localization case, the analysis by small determinants is expected to provide good approximated value for the localization exponent $\nu$. Our obtained result from bootstrap is $\nu= 2.5-2.75$ in the two dimensional case, and $\nu= 1.01$ for three dimensional case, which are obtained $4\times 4$ determinants. The numerical result by the finite size scaling for $\nu$ in the two dimensions is 2.75, and the lower critical dimension is estimated as $D_c= 1.44-1.55$ \cite{Ueoka2017}. The difference 
  between of the finite size scaling results and the bootstrap method will be expected to be small by the increasing  size of the determinants in taking more scaling dimensions $\Delta$.

  \vskip 3mm
   \section{Conformal bootstrap}
   \vskip 3mm
  
The briefly explanation of  the determinant method for the conformal bootstrap theory is following \cite{Gliozzi2013}.  The conformal bootstrap theory  is based on the conformal group $O(D,2)$, and the conformal block $G_{\Delta,L}$ is the eigenfunction of the second Casimir differential operator $ \tilde D_2$.
The eigenvalue of this  Casimir operator is denoted by $C_2$, 
\ba\label{Casimir}
&&\tilde D_2 G_{\Delta,L} = C_2 G_{\Delta,L},\nonumber\\
&&C_2 = \frac{1}{2}[\Delta (\Delta - D) + L( L + D - 2)].
\ea
where $\Delta$ is a scale dimension, which represents  the index of the singularity, and $D$ is a space dimension, and we take it as a positive continuous real number.  $L$ is a spin, related to the angular momentum, and takes an even integer
value $L=0,2,4,...$. We call $L$ as 
higher spin if the value of $L$ is large. 
The solutions of the Casimir equation have been studied  in many places, for instance in \cite{James1968,Koornwinder1978}.  They have been investigated in detail for the conformal bootstrap in \cite{Dolan2004}. The conformal block $G_{\Delta,L}(u,v)$ has two variables $u$ and $v$, which denote the cross ratios, $u= (x_{12}x_{34}/x_{13}x_{24})^2$ , $v=(x_{14}x_{23}/x_{13}x_{24})^2$, where $x_{ij}= x_i-x_j$ ($x_i$ is complex  coordinate). They  are expressed as $u= z\bar z$ and $v=(1-z)(1-\bar z)$.
For the particular point $z=\bar z$ (symmetrical point),  the conformal block $G_{\Delta,L}(u,v)$ for spin zero (L=0) case  has a simple expression by the hypergeometric function,
\be\label{hyper}
G_{\Delta,0}(u,v)|_{z=\bar z} = {(\frac{z^2}{1-z})^{\Delta/2}}  { }_3F_2[\frac{\Delta}{2},\frac{\Delta}{2},\frac{\Delta}{2}-\frac{D}{2}+1;\frac{D+1}{2},\Delta-\frac{D}{2}+1; \frac{z^2}{4(z-1)}]
\ee
The conformal bootstrap means that the  scaling dimensions $\Delta$ are determined by  the condition of the equivalence of the crossing symmetry $x_1\leftrightarrow x_3$.
For the purpose of practical calculations, the point $z=\bar z = 1/2$ is chosen usually, and we take this symmetrical point and consider the Taylor expansion at this symmetric point in this paper.    The conformal bootstrap  analysis by a small size of matrix  has been investigated for Yang-Lee edge singularity with   rather accurate results 
of the scale dimensions by Gliozzi \cite{Gliozzi2013}. In the paper of  \cite{Hikami2018a}, further the minors of conformal block for Yang-Lee edge singularity were investigated by the Pl\"ucker relations. 

There are infinite numbers of the scaling dimensions $\Delta$ and $L$, and we have to take all these values in the conformal bootstrap method.
The determinant method uses the  finite numbers of $\Delta$ and $L$ as an approximation. The fundamental scaling dimensions $\Delta$, which are smaller than other scaling dimensions, are $\Delta_\phi$, $\Delta_\epsilon$, which are related to the critical exponents $\eta$ and $\nu$. There are also a scaling dimension $\Delta_T$ which is related to the crossover exponent
$\varphi$. Also the scaling dimension which is related to the critical exponent $\omega$ of the correction to the scaling. The case of the value of the spin $L=0$ is  scalar, and 
$L=2$ case corresponds energy momentum tensor. $L=4$ represents the fourth derivatives , and this 4-spin is denoted as Q (quartic) in this paper. 
We consider in this paper up to the order of 6 derivatives (L=6, sextic), and neglecting more higher order spins $L$. 

For the replica limit of $N\to 0$ as explained in the previous section, we have a condition $\Delta_T=\Delta_\epsilon$. The Anderson localization has a constraint $\Delta_\phi=0$. Therefore,  the value of $\Delta_\epsilon=\Delta_T$ is determined minimumly from the
intersection point of the roci of the minors of $3\times3$ matrices, whose elements involve $\Delta_\epsilon,\Delta_T$ and the scaling dimension of energy momentum tensor of $L=2$(equal to $D$). 

 The bootstrap method is comprised of the crossing symmetry of the four point function.
The  four point correlation function for the scalar field $\phi(x)$ is given by
\be\label{eq1}
<\phi(x_1)\phi(x_2)\phi(x_3)\phi(x_4)> = \frac{g(u,v)}{|x_{12}|^{2\Delta_\phi}|x_{34}|^{2\Delta_\phi}}.
\ee
The notation of $x_{ij}=x_i-x_j $ ($x_i$ is complex coordinate) and the scaling dimension $\Delta_\phi$  for the field $\phi(x)$ are used. The amplitude  $g(u,v)$ is expanded as the sum of conformal blocks $G_{\Delta,L}$ ($L$ is a spin),
\be\label{eq2}
g(u,v) = 1 + \sum_{\Delta,L} p_{\Delta,L} G_{\Delta,L}(u,v)
\ee
where $u$ and $v$ are cross ratios, defined by $u= (x_{12}x_{34}/x_{13}x_{24})^2$ and $v= (x_{14}x_{23}/x_{13}x_{24})^2$.

The crossing symmetry of the exchange $x_1\leftrightarrow x_3$ implies from (\ref{eq1}) and (\ref{eq2}),
\be\label{crossing}
\sum_{\Delta,L} p_{\Delta,L} \frac{v^{\Delta_\phi}G_{\Delta,L}(u,v)- u^{\Delta_\phi}G_{\Delta,L}(v,u)}{u^{\Delta_\phi}-v^{\Delta_\phi}} = 1.
\ee

A minor method is consist of the derivatives at the symmetric point $z=\bar z= 1/2$. 
By the change of variables $z=(a+ \sqrt{b})/2$, $\bar z= (a-\sqrt{b})/2$, derivatives  are taken about $a$ and $b$. 
Since the numbers of equations  become larger than the numbers of the  scale dimensions $\Delta$, which are truncated in the sum of (\ref{eq2}),
we need to consider the zeros of minors for the determination of the values of $\Delta$. The matrix elements of minors are expressed by,
\be\label{bootstrap}
f_{\Delta,L}^{(m,n)}= (\partial_a^m \partial_b^n \frac{v^{\Delta_\phi}G_{\Delta,L}(u,v)- u^{\Delta_\phi}G_{\Delta,L}(v,u)}{u^{\Delta_\phi}-v^{\Delta_\phi}})|_{a=1,b=0}
\ee
We have one inhomogeneous equation as
\be
\sum_{\Delta,L} p_{\Delta,L} f^{(0,0)}_{{\Delta,L}}=1
\ee
and infinite number homogeneous equations,
\be
\sum_{\Delta,L} p_{\Delta,L} f_{\Delta,L}^{(2m,n)} = 0.
\ee
To have nontrivial solution for the values of the scaling dimensions $\Delta$, it is required that the determinant made of $f_{\Delta,L}^{(2m,n)}$ should be null.
Since the numbers of the equations exceed the numbers of the scaling dimensions $\Delta$, we nee the condition such that the minors are vanishing.
The $3\times 3$ minors  $d_{ijk}$ are  the determinants such that
\be\label{dijk}
d_{ijk} = {\rm det} ( f_{\Delta,L}^{(m,n)} )
\ee
where $i,j,k$ are numbers chosen differently from  (1,...,6), following the dictionary correspondence to $(m,n)$ as  $1\to (2,0)$, $2\to (4,0)$, $3\to (0,1)$, $4\to (0,2)$,  $5\to (2,1)$, $7\to (6,0)$ and $8\to  (4,1)$ \cite{Gliozzi2013}. 
 Since we consider the scaling dimension $\Delta_T$, three components $f_{\Delta,L}$ of the $3\times 3$ determinant are  $\Delta=\Delta_\epsilon$, $\Delta=\Delta_T$ and the scaling dimension of the energy momentum $\Delta=D, (L=2)$.
 The minor $d_{123}$ is considered to be basic one, which involves the second, fourth derivatives of $a$ and the first derivative of $b$. We take account of 5 minors;  $d_{123}$, $d_{124}$, $d_{125}$, $d_{234}$ and $d_{134}$.
 The minor $d_{123}$  is expressed explicitly,
\be
d_{123}= {\rm det} \left(\begin{array}{rrr} f_{\Delta_\epsilon,L=0}^{(2,0)} & f_{(D,2)}^{(2,0)}  & f_{\Delta_T,L=0}^{(2,0)}\\
f_{\Delta_\epsilon,L=0}^{(4,0)} & f_{(D,2)}^{(4,0)} & f_{\Delta_T,L=0}^{(4,0)}\\
f_{\Delta_\epsilon,L=0}^{(0,1)} & f_{(D,2)}^{(0,1)}  & f_{\Delta_T,L=0}^{(0,1)}\end{array} \right)
\ee

 We have Pl\"ucker relation \cite{Hikami2018a,Bruns} for a $3\times 6$ matrix such that
 \be\label{3pl}
[146][235]+ [124][356]-[134][256]+[126][345]-[136][245]+[123][456]=0
\ee
where we use the notation of $[ijk]= d_{ijk}$. Also we have
\be\label{tableau}
[146][235]= - 3 [123][456]-[125][346]+[135][246]
\ee
Thus the roci of zeros of these minors are not independent.
 \vskip 3mm
 
 \section{$3\times 3$ minors}
 \vskip 3mm
 
 We take $\Delta_\phi=0$ for the localization problem as (\ref{condition}). The loci of the minors $d_{ijk}=0$ are shown in $(\Delta_\epsilon,\Delta_T)$ plane. We neglect other $\Delta$ and spin $L > 2$ as a first approximation. There are no unknown parameters except $\Delta_\epsilon,\Delta_T$ in  $3\times 3$ minors. Thus, The value of $\Delta_\epsilon$ is 
 determined completely. The scaling dimension of the energy momentum tensor is fixed to $\Delta=D$. For $4\times 4$ minors, for instance, we have to introduce one parameter of the scaling dimension $\Delta_4$ for quartic spin (4-spin) and the results of the analysis may depend on the value of $\Delta_4$. In $3\times 3$ minors, there is no such ambiguity. 
 
 \vskip 3mm
For $D=2.0$, the zero loci of $d_{123},d_{124},d_{134},d_{125}$ are shown in Fig.1. The value $\Delta_\epsilon=1.52$ obtained from $d_{123}$ is different from the numerical analysis of
the finite size scaling, which provides $\Delta_\epsilon= 1.634$ \cite{Asada2002}.

\begin{figure}
{\bf{Fig.1}}
\hskip 19mm {$\Delta_T$}
\vskip 2mm
\centerline{\includegraphics[width=0.5\textwidth]{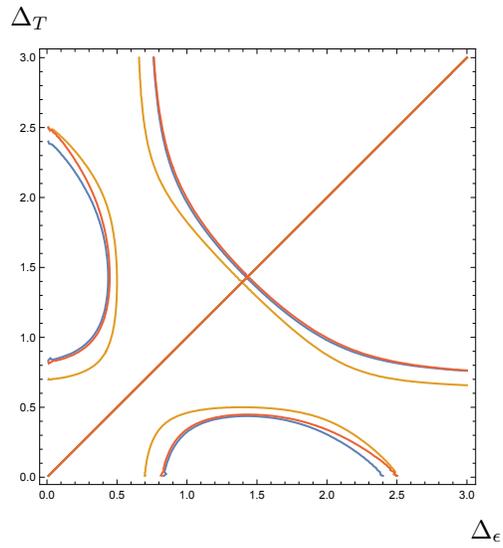}}
\vskip 1mm \hskip 89mm { $\Delta_\epsilon$}
\caption{ Localization in D=2 : The fixed point (intersection point) $\Delta_\epsilon$ = $\Delta_T = 1.52$   is obtained.} 
\end{figure}
\vskip 2mm

For $D=3$, the zero loci of $d_{123},d_{124},d_{134},d_{125}$ are shown in Fig.2.
\begin{figure}
{\bf{Fig.2}}
\hskip 19mm {$\Delta_T$}
\vskip 2mm
 \centerline{\includegraphics[width=0.5\textwidth]{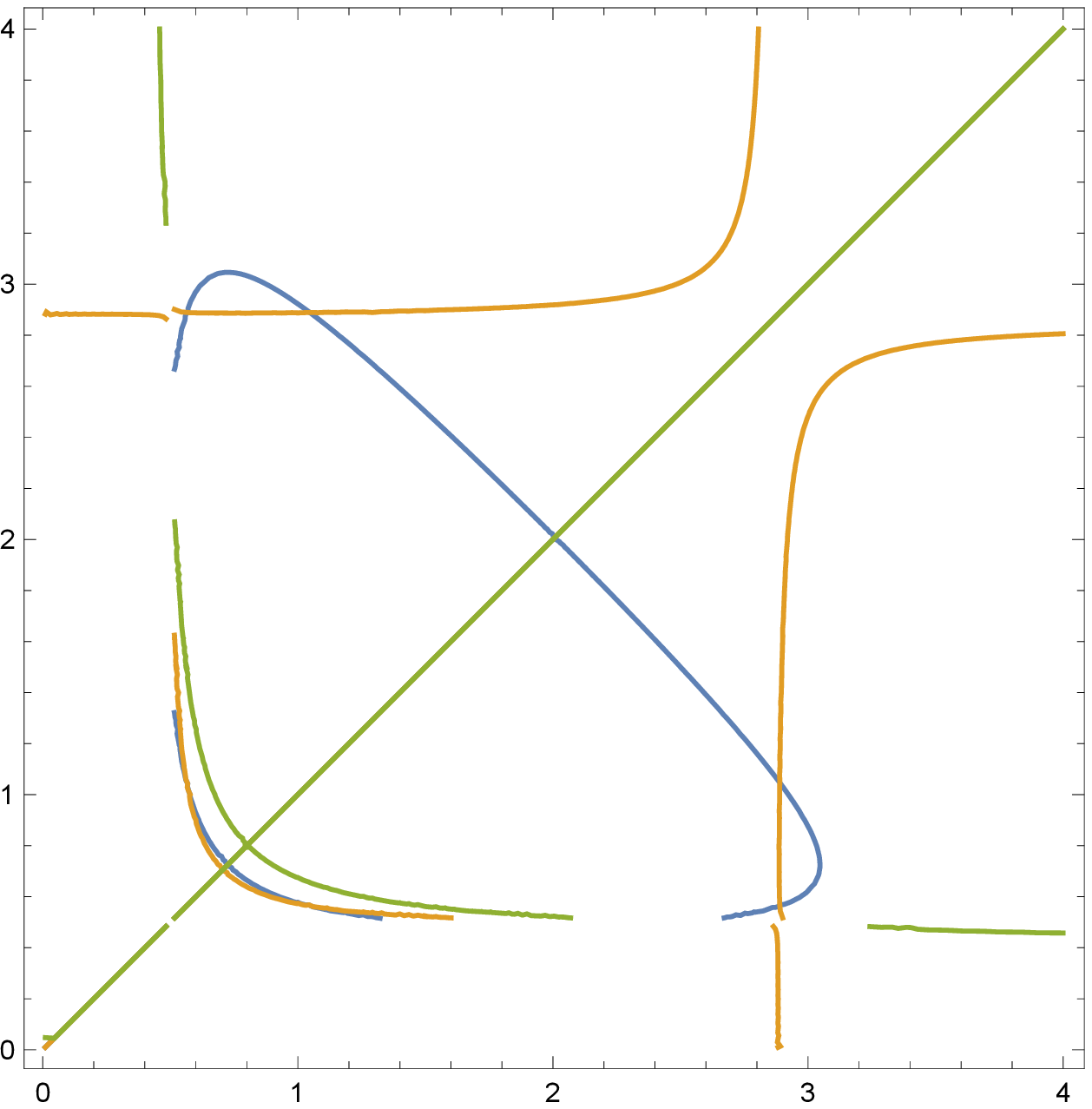}}
 \vskip 1mm \hskip 89mm { $\Delta_\epsilon$}
\caption{ Localization in D=3 : The fixed point $\Delta_\epsilon$ = $\Delta_T = 2.01$   is obtained from the zero loci of $d_{123}$. }
\end{figure}
\vskip 2mm
\vskip 3mm

For $D=1.5$, the zero loci of $d_{123},d_{124},d_{134},d_{125}$ are shown in Fig.3.
\begin{figure}
{\bf{Fig.3}}
\hskip 19mm {$\Delta_T$}
\vskip 2mm
 \centerline{\includegraphics[width=0.5\textwidth]{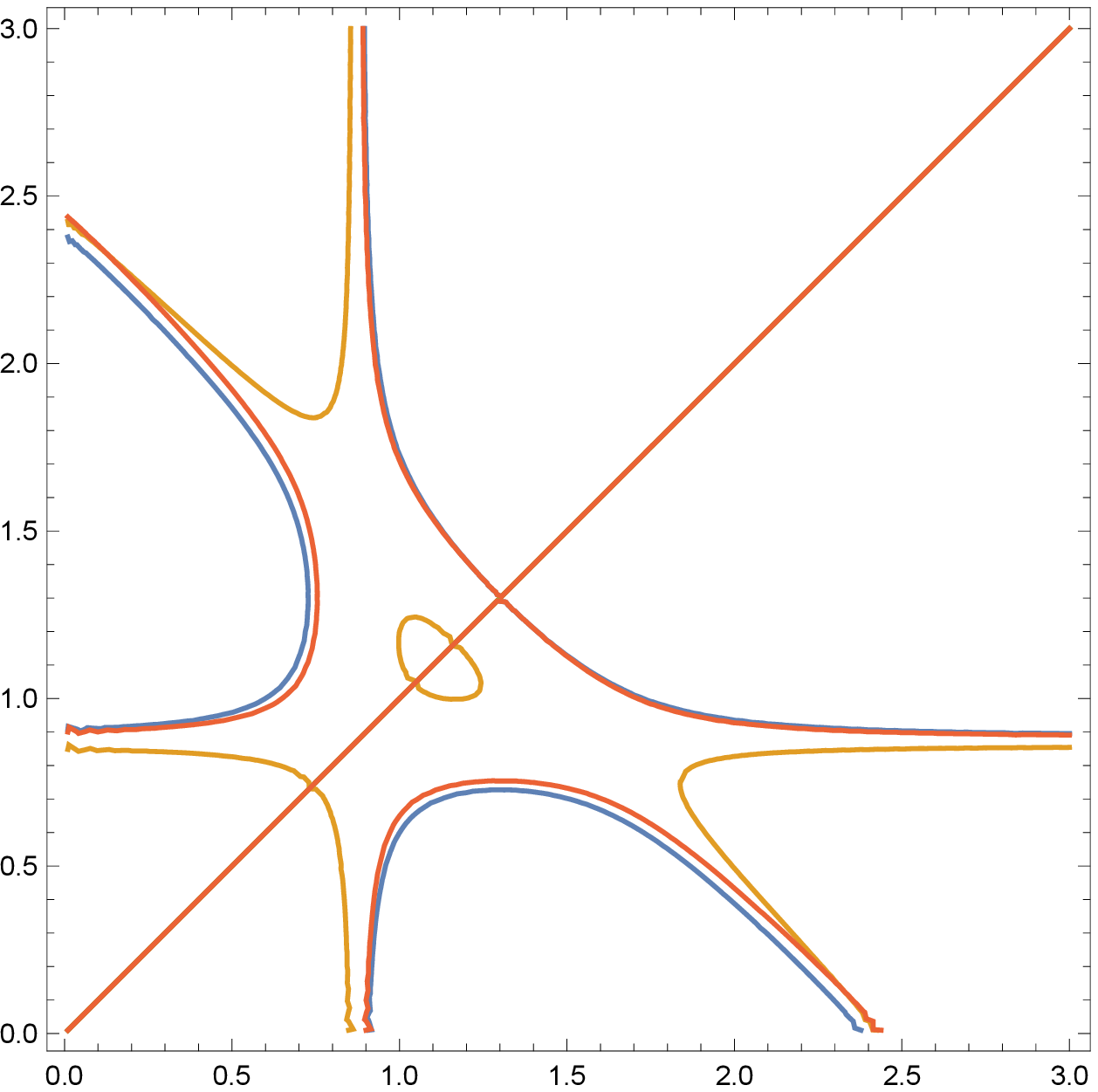}}
 \vskip 1mm \hskip 89mm { $\Delta_\epsilon$}
\caption{ Localization in D=1.5 : The fixed point $\Delta_\epsilon$ = $\Delta_T = 1.30$   is obtained. 
 }.
\end{figure}
\vskip 2mm
\begin{figure}
{\bf{Fig.4}}
\hskip 19mm {$\Delta_T$}
\vskip 2mm
 \centerline{\includegraphics[width=0.5\textwidth]{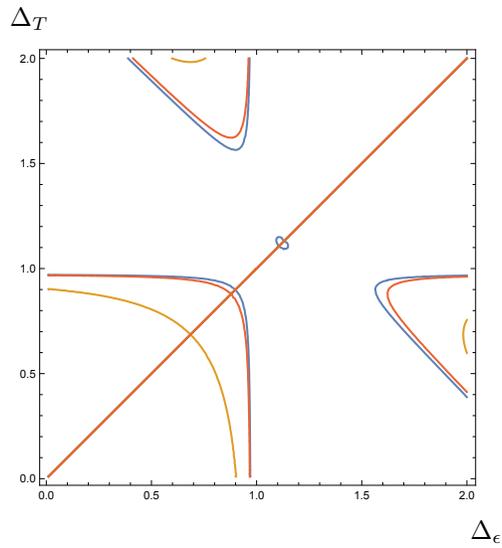}}
  \vskip 1mm \hskip 89mm { $\Delta_\epsilon$}
\caption{ Localization in D=1.247 from $3\times 3$ determinant $d_{123}$: The fixed point $\Delta_\epsilon$ = $\Delta_T = 1.12$   is obtained from the central circle, which  disappears at $D=1.245$.}
\end{figure}
\vskip 45mm

{\bf{Table 1.}}
{\bf{Scale dimension $\Delta_\epsilon$  from $3\times 3$ determinant $d123$}}. 
The value of $\Delta_\epsilon = \Delta_T$ is obtained from the intersection point of the loci of $d_{123}=0$ and the line of $\Delta_\epsilon=\Delta_T$. 
For $D=1.245$, there is no solution as indicated by (*). The value of $\nu$ becomes smaller than the mean field value $0.5$ for $D > 6$, which means the fixed point
 may be unstable in this $3\times 3$ determinant method.

\begin{tabular}[t]{|c|c|c|c|}
\hline
 $D $ & $\Delta_T=\Delta_\epsilon$ &$\frac{1}{\nu} = D - \Delta_\epsilon$& $\nu$ \\
 \hline
 1.245 & * & * & *\\
 1.247 & 1.12 & 0.13 & 7.69\\
1.5& 1.30 &  0.20 & 5.0\\
2& 1.52 & 0.48 & 2.08\\
3 & 2.01 & 0.99 & 1.01\\
4 & 2.56& 1.44 & 0.694\\
5 & 3.15 & 1.85 & 0.541\\
6 & 3.77 & 2.23 & 0.448\\
7 & 4.40 & 2.60 & 0.385\\
8 & 5.05 &\ 2.95 & 0.339\\
10 & 6.38 & 3.62 & 0.276\\
\hline
\end{tabular}

\newpage

\section{ $4\times 4$ determinants     }

\vskip 2mm
It is easy to increase the numbers of operators. For instance $4\times 4$ or $5\times 5$ determinants, the spin 4 operator L=4, and spin 6 operator can be included, 
also including the operator for the correction to scaling.

The problem is that we don't know the values of scale dimension of spin 4 or spin 6. These values are treated as  parameters for the higher rank determinants.
 
The $4\times 4$ minor  $d_{1237}$ is written as 
\be
d_{1237}= {\rm det} \left(\begin{array} {rrrr} f_{\Delta_\epsilon,L=0}^{(2,0)} & f_{(D,2)}^{(2,0)}  & f_{\Delta_T,L=0}^{(2,0)} & f_{Q}^{(2,0)}\\
f_{\Delta_\epsilon,L=0}^{(4,0)} & f_{(D,2)}^{(4,0)} & f_{\Delta_T,L=0}^{(4,0)} & f_{Q}^{(4,0)}\\
f_{\Delta_\epsilon,L=0}^{(0,1)} & f_{(D,2)}^{(0,1)}  & f_{\Delta_T,L=0}^{(0,1)} & f_{Q}^{(0,1)}\\
f_{\Delta_\epsilon,L=0}^{(6,0)} & f_{(D,2)}^{(6,0)}  & f_{\Delta_T,L=0}^{(6,0)} & f_{Q}^{(6,0)}\\\end{array} \right)
\ee
The indices of $d_{ijkl}$ corresponds to the dictionary rule; $1\to (2,0)$, $2\to (4,0)$, $3\to (0,1)$, $4\to (0,2)$,  $5\to (2,1)$, $7\to (6,0)$ and $8\to  (4,1)$. 

 For $D=2.0$, from $d_{1235}=0$, we obtain the intersection points $\Delta_\epsilon = 1.635$,
 by putting the value of $Q$ as $Q=1.55$ as Fig. 5. The value $\Delta_\epsilon = 1.635$ leads to $\nu = 2.74$.
 In Fig.5, the zero loci of four minors, $d_{1235}$,$d_{1237}$,$d_{1357}$ and $d_{2347}$ are shown. The two lines of $d_{1235}$ and $d_{1357}$ are overlapping (degenerate) for $Q=1.55$.
 For $Q=1.7$, these four lines become separated as inFig.6, and the degeneracy between $d_{1235}$ and $d_{1357}$ is resolved for $Q=1.7$.
 
 The obtained value $\Delta_\epsilon=1.635$, which provides $\nu=2.74$ is very closed the estimate by the finite scaling analysis, which gives $\nu=2.73\pm0.02$ for the
 symplectic localization in two dimensions \cite{Asada2002}.
  
\begin{figure}
{\bf{Fig.5}}
\hskip 19mm {$\Delta_T$}
\vskip 2mm
 \centerline{\includegraphics[width=0.5\textwidth]{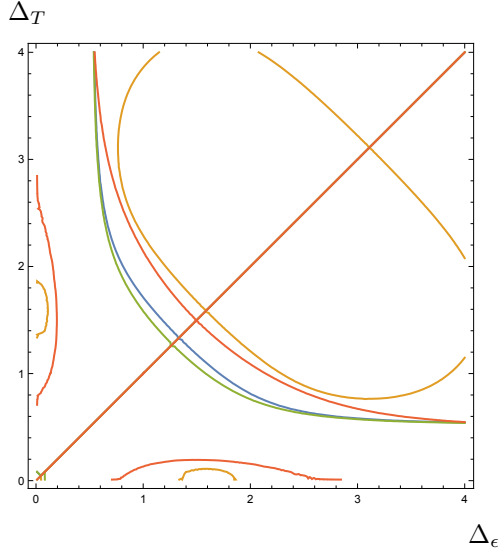}}
   \vskip 1mm \hskip 89mm { $\Delta_\epsilon$}
\caption{ Localization in D=2.0 by $4 \times 4$ determinants of $d_{1235}$,$d_{1237}$,$d_{2347}$: The fixed point $\Delta_\epsilon$ = $\Delta_T$ = $1.6$ are obtained  with the value  $Q=2.5$ from $d_{1357}$. }.
\end{figure}

\vskip 3mm
For $D=2$, if we take $Q=2.5$, there appears an elliptic circle by $d_{1357}$ which intersects the straight line of $\Delta_\epsilon=\Delta_T$ at $\Delta_\epsilon =1.6$. This leads to 
$\nu=2.5$. Other loci do not show a circle shape. The point at which $d_{1357}$ does not move so much when $Q$ is changed. For instance if we take $Q=1.55$, we obtain
the estimation $\Delta_\epsilon=1.635$, which leads to $\nu =2.74$.
 \vskip 3mm
 \begin{figure}
 {\bf{Fig.6}}
 \hskip 19mm {$\Delta_T$}
 \vskip 2mm
 \centerline{\includegraphics[width=0.5\textwidth]{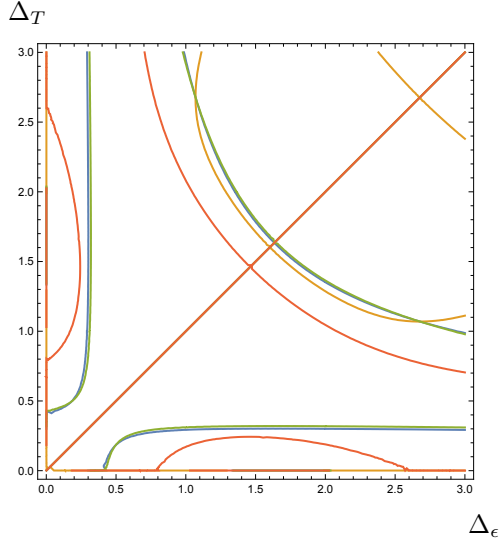}}
   \vskip 1mm \hskip 89mm { $\Delta_\epsilon$}
\caption{ Localization in D=2.0 by $4 \times 4$ determinant $d_{1235}$,$d_{1237}$,$d_{1357}$ and $d_{2347}$: The fixed point $\Delta_\epsilon$ = $\Delta_T$ = $1.635$ are obtained from
$d_{1234}$ and $d_{1357}$ with the value  $Q=1.55$. This value leads to $\nu=2.74$.}
\end{figure}

  \vskip 3mm
%  \begin{figure}
%   {\bf{Fig.7}}
% \hskip 19mm {$\Delta_T$}
% \vskip 2mm
% \centerline{\includegraphics[width=0.5\textwidth]{d1235Q=255D2p5.eps}}
 %  \vskip 1mm \hskip 89mm { $\Delta_\epsilon$}
%\caption{ Localization in D=2.5 by $4 \times 4$ determinants of $d_{1235}$,$d_{1237}$, $d_{1357}$, $d_{2347}$: The fixed point $\Delta_\epsilon$ = $\Delta_T$ = $1.68$ are obtained  with the value  $Q=2.55$.  }
%\end{figure}
%\vskip 3mm

For $D=2.5$, $\Delta_\epsilon=\Delta_T= 1.68$ is obtained for $Q=2.55$.

For $D=3$, the value of $\Delta_\epsilon$ is estimated as 2.05 by the degeneracy of $d_{1235}$,$d_{1357}$ and $d_{2347}$ as shown in Fig. 7, where $Q=3.2$ is taken. 
This leads to $\nu= 1.05$, which is a bit less than the numerical estimate of $\nu= 1.3$ \cite{Asada2002}. If the value $Q$ decreases, the value of $\nu$ increases from the
intersection of $d_{1235}$.  
\vskip 3mm
\begin{figure}
   {\bf{Fig.7}}
 \hskip 19mm {$\Delta_T$}
 \vskip 2mm
 \centerline{\includegraphics[width=0.5\textwidth]{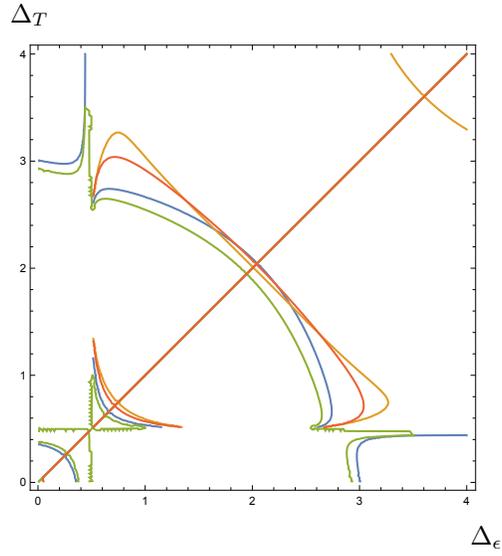}}
   \vskip 1mm \hskip 89mm { $\Delta_\epsilon$}
\caption{ Localization in D=3.0 by $4 \times 4$ determinants of $d_{1235}$,$d_{1237}$,$d_{2347}$: The fixed point $\Delta_\epsilon$ = $\Delta_T$ = $2.05$ are obtained  with the value  $Q=3.2$. }
\end{figure}
\vskip 3mm

For $D=4$, as shown in Fig.8, the four lines of zero loci of the determinants $d_{1235}$, $d_{1237}$, $d_{1357}$ and $d_{2347}$ intersect the  line of $\Delta_\epsilon$ = $\Delta_T$
at $\Delta_\epsilon = 2.93$, which leads to the value of $\nu=0.935$.
\vskip 3mm
\begin{figure}
   {\bf{Fig.8}}
 \hskip 19mm {$\Delta_T$}
 \vskip 2mm
 \centerline{\includegraphics[width=0.5\textwidth]{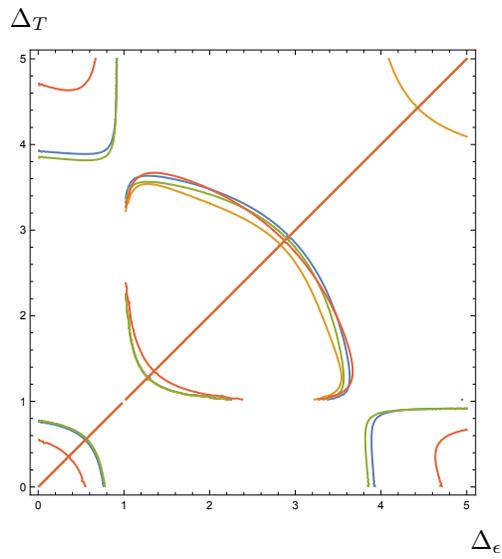}}
   \vskip 1mm \hskip 89mm { $\Delta_\epsilon$}
\caption{ Localization in D=4.0 by $4 \times 4$ determinants $d_{1235}$, $d_{1237}$, $d_{1357}$ and $d_{2347}$: The fixed point $\Delta_\epsilon$ = $\Delta_T$ = $2.93$ are obtained  with the value  $Q=4.2$}.
\end{figure}

%\vskip 3mm
%\begin{figure}
%   {\bf{Fig.10}}
% \hskip 19mm {$\Delta_T$}
% \vskip 2mm
% \centerline{\includegraphics[width=0.5\textwidth]{d1235Q=525D5p0.eps}}
%    \vskip 1mm \hskip 89mm { $\Delta_\epsilon$}
%\caption{ Localization in D=5.0 by $4 \times 4$ determinants $d_{1235}$, $d_{1237}$, $d_{1357}$ and $d_{2347}$: The fixed point $\Delta_\epsilon$ = $\Delta_T$ = $3.75$ are obtained  with the value  $Q=5.25$. }
%\end{figure}
%\vskip 3mm
%\begin{figure}
%   {\bf{Fig.11}}
 %\hskip 19mm {$\Delta_T$}
 %\vskip 2mm
 %\centerline{\includegraphics[width=0.5\textwidth]{d1235Q=635D6p0.eps}}
%   \vskip 1mm \hskip 89mm { $\Delta_\epsilon$}
%\caption{ Localization in D=6.0 by $4 \times 4$ determinants $d1235$, $d1237$, $d1357$ and $d2347$: The fixed point $\Delta_\epsilon$ = $\Delta_T$ = $4.5$ are obtained  with the value  $Q=6.35$. }
%\end{figure}

\vskip 3mm
\begin{figure}
   {\bf{Fig.9}}
 \hskip 19mm {$\Delta_T$}
 \vskip 2mm
 \centerline{\includegraphics[width=0.5\textwidth]{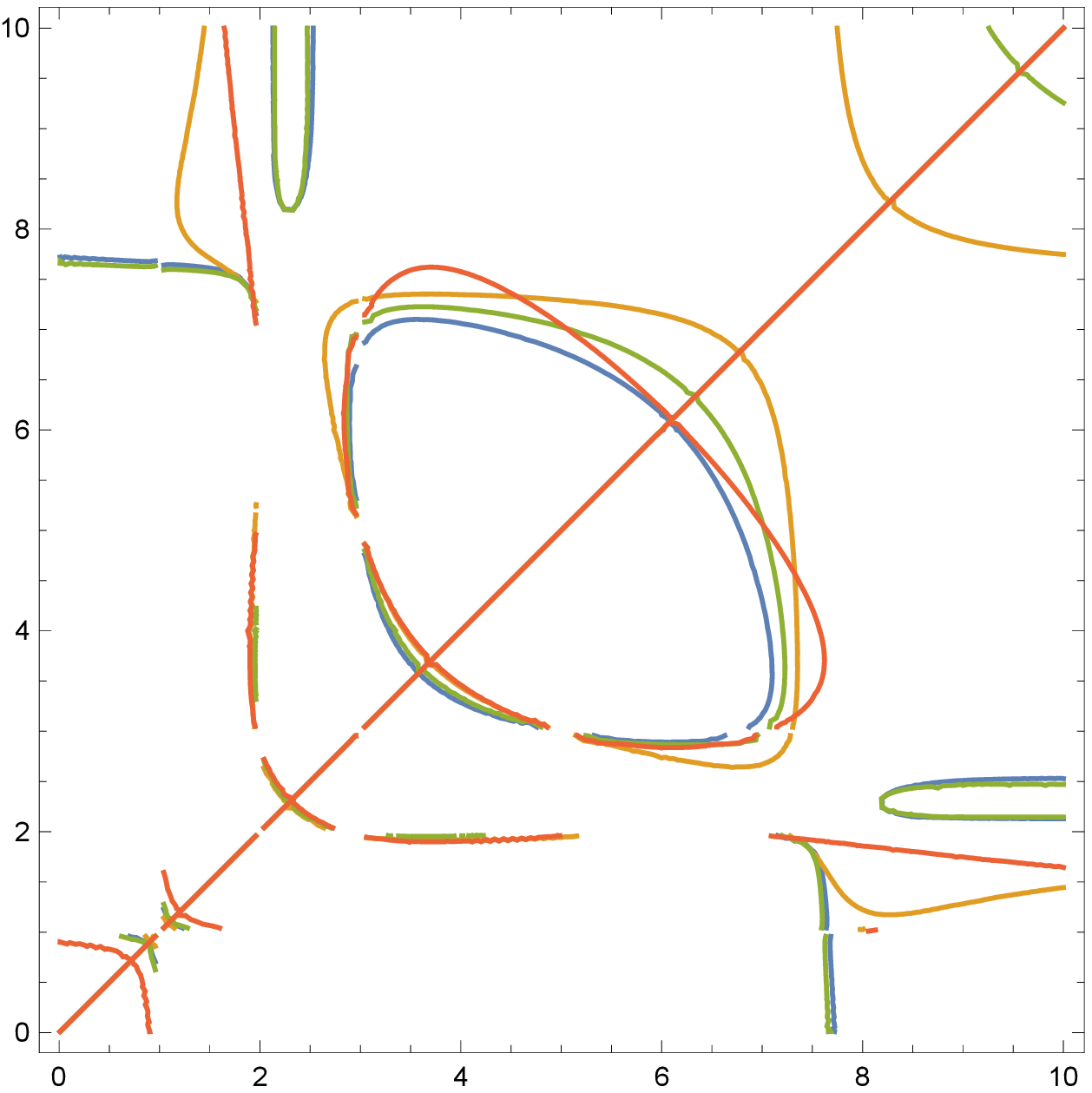}}
   \vskip 1mm \hskip 89mm { $\Delta_\epsilon$}
\caption{ Localization in D=8.0 by $4 \times 4$ determinants $d_{1235}$, $d_{1237}$, $d_{1357}$ and $d_{2347}$: The fixed point $\Delta_\epsilon$ = $\Delta_T$ = $6.1$ are obtained  with the value  $Q=8.45$. }
\end{figure}

\vskip 80mm

We present the values of the exponent $\nu$ obtained from the intersection point in Table 2.
\vskip 2mm
{\bf Table 2. Estimation of $\nu$ by $4\times 4$ determinants for various dimensions}
\vskip 2mm
\begin{tabular}[t]{|c|c|c|c|}
\hline
 $D $ & Q & $\Delta_T=\Delta_\epsilon$  & $\nu$ \\
 \hline
 2& 2.5 & 1.6  & 2.5 \\
 2.5 & 2.55 & 1.68 & 1.22 \\
3 &   3.2 & 2.05 & 1.05 \\
4 & 4.2 & 2.93 & 0.93\\
5 & 5.25 & 3.75  & 0.80\\
6 & 6.35 & 4.5 & 0.66\\
8 & 8.45 & 6.1 & 0.52 \\

\hline
\end{tabular}

\newpage

\section{Discussion}

In this paper, we examine the conformal bootstrap method for the localization by putting $\Delta_\phi=0$ condition in the blow up plane
where $\Delta_\epsilon$ differs slightly from $\Delta_T$. Similar analysis has been done for the replica $N\to 0$ limit of polymer, branched polymer and for the random field Ising model \cite{Hikami2018a,Hikami2018b,Hikami2018c}.

There is a fixed point as shown in Table 2 in the region of $2 < D < 8$, and the value of the critical exponent $\nu$ is similar to the  values by the finite size scaling
for $D=2$ and $D=3$ \cite{Minakuchi,Asada2002}. This case is symplectic localization class, where the phase transition exists in two dimensions. We are not yet conclusive
about the upper critical dimension. Around $D=8$, we have no clear fixed point. Usually the upper critical point becomes a free field, in which almost all zero loci of minors
intersect at a single point, similar to polymer and branched polymers \cite{Hikami2018b}. At $D=8$, we have no such phenomena, and this suggests that $D=8$ may not be the upper critical point
of the localization. The upper critical dimension $D=\infty$ has been suggested for the localization \cite{Ueoka2017,Biroli2017}, and our present work does not exclude this possibility.

The extension of this article is possible for including the spin 6 operators and also including the scaling dimensions such as  the exponent of the  correction of  scaling. Some calculations of these extensions do not change the present results so much. The other localization
of the orthogonal and unitary classes are remained in a future work.

The result obtained here by the simple conformal bootstrap suggests  the relevance to the symplectic Anderson localization of the strong spin orbit coupling \cite{Minakuchi}. There are possibilities 
that the present analysis  corresponds to other universality class, such as quantum Hall effect or as sated in the introduction, such as $Sp(N)/U(N)$ case. This will remain an interesting study.

 \vskip 3mm
 {\bf Acknowledgement}
 \vskip 3mm
 The author thanks Ferdinando Gliozzi and Edouard Br\'ezin for  discussions of the determinant method. Part of this work was presented in the workshop of BMFT in Rome
 University in January 3rd, 2018 and author thanks Georgi Parisi for this invitation. He thanks Adam Nahum for the comment of the exponent of quantum Hall effect, which is similar to Anderson localization. He also thank Wolfram Research  for the numerical evaluations by Mathematica 11.
 This work is supported by   JSPS KAKENHI  Grant-in Aid (C) 16K05491 and Grant-in Aid (B) 19H01813.

  \newpage
%%%%%%%%%%%%%%%%%%%%%%

\end{document}